\documentclass[12pt,reqno,a4paper]{amsart}
\usepackage{amsmath,amssymb,dsfont,graphicx,cite,rotating,setspace}
\def\beq{\begin{equation}}
\def\eeq{\end{equation}}
\def\bea{\begin{eqnarray}}
\def\eea{\end{eqnarray}}
\newtheorem{theorem}{Theorem}

\makeatletter
\expandafter\let\expandafter
\reset@font\csname reset@font\endcsname
\def\subeqnarray{\arraycolsep1pt
    \def\@eqnnum\stepcounter##1{\stepcounter{subequation}
        {\reset@font\rm(\theequation\alph{subequation})}}
\jot5mm     \eqnarray}

\makeatother

\newcommand{\CS}{{\mathcal S}}

\newcommand{\CT}{{\mathcal T}}

\def\ri{{\rm{i}}}

\def\su2{{\mathfrak {su}}(2)}
\def\e3{{\mathfrak {e}}(3)}

\begin{document}
\title[Multiscale expansion and integrability properties of the lpKdV]
{\bf Multiscale expansion and integrability properties of the lattice potential KdV equation}
\author[R. HERNANDEZ HEREDERO, D. LEVI, M. PETRERA and C. SCIMITERNA]
{R. HERNANDEZ HEREDERO${}^\dag$, D. LEVI${}^\diamond$, M. PETRERA${}^{\sharp,\diamond}$ and C. SCIMITERNA${}^{\flat,\diamond}$}

\maketitle

\centerline{\it ${}^\dag$Departamento de Matem\'atica Aplicada,} 
\centerline{\it Escuela Universitaria de Ingenier\'ia T\'ecnica de Telecomunicaci\'on,}
\centerline{\it
Universidad Polit\'ecnica de Madrid (UPM),}
\centerline{\it Campus Sur Ctra de Valencia Km. 728031, Madrid, Spain}
\centerline{e-mail: \texttt{rafahh@euitt.upm.es}}

\vspace{.3truecm}

\centerline{\it ${}^\diamond$Dipartimento di Ingegneria Elettronica,}
\centerline{\it Universit\`a degli Studi Roma Tre and Sezione INFN, Roma Tre,}
\centerline{\it Via della Vasca Navale 84, 00146 Roma, Italy}
\centerline{e-mail: \texttt{levi@fis.uniroma3.it}}

\vspace{.3truecm}

\centerline{\it ${}^\sharp$Zentrum Mathematik,}
\centerline{\it Technische Universit\"at M\"unchen,}
\centerline{\it Boltzmannstr. 3, D-85747 Garching bei M\"unchen, Germany}
\centerline{e-mail: \texttt{petrera@ma.tum.de}}

\vspace{.3truecm}

\centerline{\it ${}^\flat$Dipartimento di Fisica e Dipartimento di Ingegneria Elettronica,}
\centerline{\it Universit\`a degli Studi Roma Tre and Sezione INFN, Roma Tre}
\centerline{\it Via della Vasca Navale 84, 00146 Roma, Italy}
\centerline{e-mail: \texttt{scimiterna@fis.uniroma3.it}}

\begin{abstract}
\noindent
We apply the discrete multiscale expansion  to the Lax pair and to the first few symmetries of the lattice potential Korteweg-de Vries equation. From these calculations we show that, like the lowest order secularity conditions 
give a nonlinear Schr\"odinger equation,  the Lax pair gives at the same order the Zakharov and Shabat spectral problem and the symmetries the hierarchy of point and generalized symmetries of the nonlinear Schr\"odinger equation.

\end{abstract}

\section{Introduction}
Reductive perturbation techniques \cite{t1,t2} have proved to be important tools for finding approximate solutions of many
physical problems, by reducing a given nonlinear partial differential
equation to a simpler equation,  often integrable~\cite{ce}, and for proving
integrability~\cite{ce,KZ,dms,km,dp}. Recently, after various attempts to
carry over this approach  to partial difference equations \cite{Ag,lm,LevHer} we have presented a procedure for carrying out a multiscale expansion on the lattice \cite{levi,lp,HLPS} which seems to preserve the integrability properties \cite{HLPS_t}.
To get a better understanding of the application of the reductive perturbation technique on difference equations, after an introduction in Section \ref{mel} on multiscale expansions on the lattice  potential KdV equation (lpKdV), we discuss in Section \ref{sp} its application to the spectral operator, as was done by Zakharov and Kuznetsov in their pioneering work in 1986 \cite{KZ} for the KdV equation.   Later on we apply, in Section \ref{gs}, the multiscale expansion to the symmetries of the lpKdV \cite{lp_sym}. Section \ref{c} is devoted to a few conclusive remarks.

\section{Multiscale expansion on the lattice} \label{mel}

The aim of this Section is to give a terse survey on the multiscale analysis on the lattice and its application to the reduction of the lpKdV. We refer to
\cite{levi,lp,HLPS} for further details.

\subsection{Shift operators  defined on the lattice}

Let $u_n: \mathbb{Z}\rightarrow \mathbb{R}$ be a function defined on a lattice of index $n\in\mathbb{Z}$. 
One can always extend it to a function $u(x): \mathbb{R} \rightarrow \mathbb{R}$ by 
defining a real continuous variable 
$x = n \sigma_x$, where $\sigma_x \in \mathbb{R}$ is the constant lattice spacing.

An equation defined on the lattice is a functional relation between the function~$u_n$ and its
shifted values $u_{n \pm 1}$, $u_{n \pm 2}$, etc, expressed in terms of
 a shift operator $T_n$
such that~$T_n u_n = u_{n+1}$. 

For the continuous function $u(x)$ we can introduce an operator $T_x$,
such that $T_x u(x)= u(x+\sigma_x)$.
The Taylor expansion of $u(x+\sigma_x)$ centered in $x$ reads
\beq \label{tay1}
T_x u(x)= \sum_{i=0}^\infty \frac{\sigma_x^i}{i!} u^{(i)}(x),
\eeq
where $u^{(i)}(x) = d^i u(x) /dx^i = d_x^i u(x)$, with $d_x$  the total derivative. Eq.~(\ref{tay1}) 
suggests  the following formal expansion for the differential operator $T_x$:
\beq \nonumber
T_{x}= e^{\sigma_x d_{x}}= \sum_{i=0}^\infty \frac{\sigma_x^i}{i!} d_x^i.
\eeq
Introducing a formal derivative with respect to the index $n$, say $\delta_n$, 
we can define, by analogy with $T_x$, the operator $T_n$  as
\beq \label{tnr}
T_{n} = e^{\delta_{n}}= \sum_{i=0}^\infty \frac{\delta_n^i}{i!}.
\eeq
The formal expansion (\ref{tnr}) can be inverted, yielding
\beq \label{tnn}
\delta_{n}=\ln{T_{n}}=\ln(1+\Delta_{n})=  \sum_{i=1}^\infty \frac{(-1)^{i-1}}{i}\Delta_{n}^i,
\eeq
where $\Delta_{n}= T_{n}-1$ is the discrete right difference operator w.r.t.~the variable $n$ (i.e.~$\Delta_n u_n = u_{n+1} - u_n$).

Following~\cite{levi,lp} we say that $u_n$ is a {\it slow-varying  function of order $\ell$} iff $\Delta_n^{\ell+1}  u_n=0$.
Hence the $\delta_n$
operators are formal series containing infinite
powers of $\Delta_n$, but, acting on slow-varying functions of order $\ell$, they reduce
to polynomials in $\Delta_n$ of order at most $\ell$.

\subsection{Dilations on the lattice}

Let us introduce a second lattice, obtained from the first by 
a dilation. For $x \in\mathbb{R}$ we can  visualize the problem as a
change of variable between $x$   and  $x_1= {\epsilon}{x}$, $0 < \epsilon \ll 1$. 
On the lattice this corresponds to a change from the index $n= x/ \sigma_{x}$ 
to the new index $n_1 = x_1/ \sigma_{x_1}$, 
where $\sigma_{x_1}$ is the new lattice spacing.
 Assuming that $\sigma_{x_1} \gg \sigma_x $ we
can set  $\sigma_x = \varepsilon \sigma_{x_1}$, $0 < \varepsilon \ll 1$, so that $n_1 = \epsilon \varepsilon n$.
As $n,n_1\in \mathbb{Z}$, $\epsilon \varepsilon$ is a rational number and one can define in all generality
$\epsilon \varepsilon=  M_1/N \ll 1$ with $M_1,N\in\mathbb{N}$. However, if we want the lattice of index $n_1$ to be a sublattice of the lattice of index $n$, we have also to require that $M_1/N  =1/M$ with $M\in\mathbb{N}$.

The relation between the discrete
derivatives defined in the two lattices
is given by \cite{lp,HLPS,Jordan,LevHer}
\beq
\Delta^j_{n}  u_{n}=j! \sum_{i=j}^\infty  \frac{P_{i,j}}{ i!} \Delta^i_{n_1}  u_{n_1}.  \label{difInt}
\eeq

The coefficients
$P_{i,j}$ read
$$
P_{i,j} = \sum_{k=j}^i
 \left(\frac{M_1}{N}\right)^k \CS_{i}^{k}  \mathfrak{S}_{k}^j,
$$
where $\CS_{i}^{k}$ and $\mathfrak{S}_{k}^j$ are the Stirling numbers of the
first and second kind respectively. 

If $u_n$ is a function of infinite order of slow-varyness, i.e.~$\ell=\infty$,
then Eq.~(\ref{difInt}) implies that a finite difference 
in the discrete variable $n$ depends on an infinite number of differences on the variable 
$n_1$. 

\subsection{Discrete multiscale expansion}

Let us now consider $u_n = u_{n;n_1}$ 
as a function depending on a fast index $n$ 
and a slow index $n_1= n( M_1/N)$. 
At the continuous level, the total
derivative $d_{x}$ acting on functions $u(x;x_1)$ is the sum of partial
derivatives, i.e. $d_{x}=\partial_{x} +\epsilon\partial_{x_1}$. 
As
\beq \label{IIa}
T_{x}= e^{\sigma_x d_{x}}=
e^{\sigma_x \partial_{x}} e^{\epsilon \sigma_x \partial_{x_1}},
\eeq
we can write the total shift operator $T_n$ as
\beq
T_n =
e^{\delta_{n}}
e^{ (M_1/N)\delta_{n_1}}=\mathcal{T}_{n} \mathcal{T}_{n_1}^{(  M_1/N)}, \label{ll}
\eeq
where the partial shift operators $\CT_n,\CT_{n_1}$, defined by
$\mathcal{T}_{n}u_{n;n_1} = u_{n+1;n_1}$ and $\mathcal{T}_{n_1}u_{n;n_1} = u_{n;n_1+1}$, are given by
$$
\mathcal{T}_{n} =  \sum_{i=0}^\infty \frac{\delta_n^i}{i!}, \qquad \qquad 
\mathcal{T}_{n_1}^{( M_1/N)}=
\sum_{i=0}^{\infty}\frac{( M_1/N)^{i}}{i!}\delta_{n_1}^i, \label{deftxi}
$$
and $\delta_{n_1}$ is given by Eq.~(\ref{tnn}) with $n$ substituted by $n_1$.

Eq.~(\ref{IIa}) can be extended to the case of $K$
slow variables $x_{i}= \epsilon^{i}x$, $1 \leq i \leq K$ . Then the action of the shift operator $T_n$  on a
function $u_{n;\{n_i\}_{i=1}^K}$ depending on both  fast and slow variables can be written in terms 
of the partial shifts $\mathcal{T}_{n},\mathcal{T}_{n_i}$ as
\beq \label{txu}
T_n = \mathcal{T}_{n}\prod_{i=1}^K \mathcal{T}_{n_{i}}^{(\epsilon_{n_i})} ,
\eeq
where the ${\epsilon_{n_i}}$'s are suitable functions of $\epsilon$ and $\varepsilon$ depending parametrically on some integer coefficients $M_i \in \mathbb{N}$, $1 \leq i \leq K$.

To carry out the multiscale expansion of the  fields appearing in partial difference equations with two independent discrete variables, one has to consider the action of the operator~(\ref{txu}) on a function depending on two fast indices~$n$ and $m$, and on  a set of $K_n+K_m $ slow variables $\{n_i\}_{i=1}^{K_n}$ and $\{m_i\}_{i=1}^{K_m}$ (we shall use the
notation 
$u_{n,m; \{n_i\}_{i=1}^{K_n},\{m_i\}_{i=1}^{K_m}}$ for such functions).
Notice that in principle it is possible to consider $K_n=K_m=\infty$.
We assume a common definition of the small parameter 
$\epsilon$ for both discrete variables $n$ and $m$  but 
we denote with $M_i$ the integers for the slow variables $n_i$ and with $\tilde M_i$ the ones for $m_i$. We have:
\beq \nonumber
\epsilon_{n_i} =  \frac{M_i}{N^{i}},  \quad 1 \leq i \leq K_n,
\qquad \qquad \epsilon_{m_i} = \frac{\tilde M_i}{N^{i}}, \quad 1 \leq i \leq K_m.
\eeq

Hereafter we shall assume $K_n=1$ and $K_m=K$.

\subsection{Multiscale expansion of the lattice potential KdV equation}

The lpKdV is given by \cite{frank}:
\beq
 [\mu(T_{n}T_{m}-1)+\zeta(T_{n}-T_{m})] u_{n,m}-
(T_{n}-T_{m})u_{n,m}(T_{n}T_{m}-1) u_{n,m}=0, \label{kdv}
\eeq
where $\mu= p-q$ and $\zeta= p+q$, and $p,q$, $p\neq q$, are two real parameters.
The linear part of Eq.~ (\ref{kdv}) has a travelling wave solution of the form
$u_{n,m}=\exp{\{\ri[ \kappa n -\omega(\kappa)m]\}}$ with
\begin{equation}\label{disp}
\omega(\kappa)=-{2}\arctan{\left(\frac{\zeta+\mu}{\zeta-\mu}\tan{\frac{\kappa }{2}}\right)}.
\end{equation}

According  to \cite{HLPS} the multiscale expansion of Eq.~(\ref{kdv}) is performed taking into account that
\beq 
\label{lk}
u_{n,m}=\sum_{\alpha\in\mathbb{Z}}\sum_{k=1}^{\infty} \frac{1}{N^k}
u^{(\alpha)}_k (n_1, \{m_i\}_{i=1}^K)
e^{ \ri \alpha(\kappa
n-\omega m)}, \qquad  \qquad u_k^{(-\alpha)}=\bar u_k^{(\alpha)}.
\eeq

The following statement, proved in \cite{HLPS}, provides the multiscale expansion of the lpKdV 
(\ref{kdv}) at the lowest orders of $1/N$.

\begin{theorem} \label{th1}
The multiscale expansion of Eq.~(\ref{kdv}) gives the following results:

\begin{enumerate}

\item $\mathcal{O}(1/N)$:

\begin{itemize}

\item $\alpha=0$: the equation is identically satisfied.

\item $\alpha=1$: one gets a linear equation identically satisfied by taking into account the dispersion relation  (\ref{disp}). 

\item $ |\alpha| \geq 2$: one gets a linear equation whose only solution is $u^{(\alpha)}_{1}=0$.

\end{itemize}

\item $\mathcal{O}(1/N^2)$:

\begin{itemize}

\item $\alpha=1$: one gets a linear equation whose solution is 
\beq
u_1^{(1)}=u_1^{(1)}(n_2,\{m_i\}_{i=2}^K), \qquad n_2 =n_1\mp m_1, \label{solu}
\eeq
 provided that
$$
M_1= \mp S  \left( \mu - \zeta  e^{\ri  \kappa }  \right), \qquad \qquad
\tilde M_1= S e^{\ri  \kappa } \frac{\zeta^2 -\mu^2}{ \mu e^{\ri  \kappa }  -\zeta}.
$$
Here $S= r \exp{(\ri \theta)}$,
with $r>0$ and $\theta= -\arctan \left[ (\zeta \sin \kappa) /(\zeta \cos\kappa -\mu) \right]$, assures that 
$M_1$ and $\tilde M_1$ are positive integers.

\item $\alpha=0$:  one gets
\beq
\delta_{n_2} u_1^{(0)}= \tau_1 |u_{1}^{(1)}|^2, \qquad \qquad
\tau_1=  \pm \frac{2  \left(1+e^{\ri \kappa} \right)^2}
{ S e^{\ri \kappa} (\mu +\zeta) \left( \mu -\zeta e^{\ri \kappa} \right) } , \nonumber
\eeq
where $u_1^{(0)}=u_1^{(0)}(n_2,\{m_i\}_{i=2}^K)$.

\item $\alpha=2$: one gets
\beq
u_2^{(2)}= \tau_2 (u_{1}^{(1)})^2, \qquad \qquad 
\tau_2= \frac{1+e^{\ri \kappa}}
{(1-e^{\ri\kappa})(\mu+\zeta)} , \nonumber
\eeq
where  $u_2^{(2)}=u_2^{(2)}(n_2,\{m_i\}_{i=2}^K)$.

\end{itemize}

\item $\mathcal{O}(1/N^3)$: 

\begin{itemize}

\item $\alpha=1$: one gets the following (defocusing) dNLS:
\beq
\ri \delta _{m_2} u_1^{(1)} = \rho_1 \delta_{n_2}^2u_1^{(1)}+ \rho_2  u_1^{(1)} |u_1^{(1)}|^2, \label{nlsr}
\eeq
where 
$$
 \rho_1 = - \frac{\mu \zeta r^2 (\zeta^2-\mu^2) \sin \kappa}
{ \tilde M_2 \left( \zeta^2 +\mu^2 -2 \zeta \mu \cos \kappa \right)},  \qquad 
\rho_2= \frac{8 \zeta \mu (\zeta -\mu) (1 + \cos \kappa)^2 \sin \kappa}
{\tilde M_2(\mu+\zeta) \left(\zeta^2+\mu^2 -2  \zeta \mu \cos \kappa  \right)^2 }. 
$$

\item $\alpha=0$: one gets
\beq
\delta_{n_{2}} u_{2}^{(0)} = \tau_1 \left(  u_{1}^{(1)} \bar u_{2}^{(1)} + \bar u_{1}^{(1)} u_{2}^{(1)} \right) -
\tau_3 \left ( \bar u_{1}^{(1)} \delta_{n_2} u_{1}^{(1)}  - u_{1}^{(1)} \delta_{n_2} \bar u_{1}^{(1)} 
\right) , \nonumber
\eeq
with
$$
\tau_3 = \frac{2 \ri \sin \kappa}{\mu+\zeta},
$$
where $u_2^{(0)}=u_2^{(0)}(n_2,\{m_i\}_{i=2}^K)$ and $u_2^{(1)}=u_2^{(1)}(n_2,\{m_i\}_{i=2}^K)$.
\item $\alpha=2$: one gets
\beq
u_{3}^{(2)}=\tau_{4} u_{1}^{(1)}(\delta_{n_{2}}u_{1}^{(1)})+2\tau_{2}u_{1}^{(1)}u_{2}^{(1)}, \qquad 
\tau_4 = \pm \frac{2Se^{\ri \kappa}(\alpha+\beta e^{\ri \kappa})} {(e^{ \ri \kappa}-1)^2(\mu+\zeta )},\nonumber
\eeq
where $u_3^{(2)}=u_3^{(2)}(n_2,\{m_i\}_{i=2}^K)$.

\end{itemize}

\end{enumerate}
\end{theorem}

We have given above just those results necessary to get 
a discrete nonlinear Schr\"odinger equation (dNLS)  as a secularity condition and its symmetries.

\section{Multiscale expansion of the lpKdV spectral problem} \label{sp}

As shown in \cite{lp} there are many forms for the linear problems associated with the lpKdV. The first to be introduced \cite{frank} is given by  first order~$2\times2$ matrix difference equations. Later on \cite{lp} it was shown that the matrix Lax pair could be easily reduced to a scalar  non-symmetric difference equation of second order, used by Boiti et.~al.~\cite{boiti}  to integrate an alternative form of the equations of the Volterra hierarchy. In \cite{lp_sym} it was moreover shown that by a Miura transformation it is possible to associate the lpKdV  with the Toda spectral problem introduced by Manakov and Flaschka \cite{mf} when the field $b_n(t)=0$.   

One could start  from any of the three linear problems delined in the previous paragraph to do the multiscale expansion. However we choose as starting spectral problem the one whose second derivative is expressed in a symmetric form, i.e.~the  discrete Schr\"odinger spectral problem used to integrate the Toda and Volterra equations.

The $n$-evolution equation of the (scalar) spectral problem of the lpKdV  (\ref{kdv}) may be written as \cite{lp_sym}:
\beq
\phi_{n-1} + a_{n} \phi_{n+1} = \mu \phi_{n}, \label{sp1}
\eeq
with
$$
a_{n}= \frac{4 p^2}{\left[ 2p - (T_n^2+1)u_{n,m} \right] \left[ 2p - (T_n+T_n^{-1})u_{n,m} \right]}.
$$
Here $\mu \in \mathbb{C}$ is the spectral parameter. 

Our aim is now to perform the multiscale expansion of Eq.~(\ref{sp1}) in order to get the 
corresponding evolution equation of the  spectral problem of the dNLS  (\ref{nlsr}).
We refer to \cite{KZ} for the continuous counterpart of this analysis.

To expand Eq.~ (\ref{sp1}) we  consider the development (\ref{lk}) for the field $u_{n,m}$, with the restriction
(\ref{solu}),
while the function
$\phi_n$ will be expanded according to the formula:
\beq
\phi_n=\sum_{\alpha \, {\rm{odd}} }\sum_{k=0}^{\infty} \frac{1}{N^{k}}
\phi^{(\alpha)}_k (n_2, \{m_i\}_{i=2}^K)
e^{ \ri \alpha(\kappa
n-\omega m)/2}, \qquad  \qquad \phi_k^{(-\alpha)}=\bar \phi_k^{(\alpha)}. \label{pd}
\eeq

At order $\mathcal{O}(1)$, the multiscale analysis of Eq.~(\ref{sp1}) suggests the following expansion for the spectral
parameter $\mu$:
\beq
\mu=2 \cos \left( \frac{\kappa}{2} \right) + \sum_{k=1}^{\infty}\frac{\mu_k}{N^k}. \label{mu}
\eeq

Taking into account Eq.~(\ref{mu}) we proceed to the order $1/N$ of the multiscale expansion of Eq.~(\ref{sp1}). We have:
\beq
\delta_{n_2} \phi^{(1)}_0+ \frac{2 u^{(1)}_1}{p} \cos^2 \left( \frac{\kappa}{2} \right) \bar \phi^{(1)}_0=
- \frac{\ri \mu_1}{2 \sin \left( \frac{\kappa}{2} \right) } \phi^{(1)}_0, \label{hg}
\eeq
for $\alpha=1$. The corresponding equation for $\alpha=-1$ is given by performing
the complex conjugation of Eq.~(\ref{hg}). The coefficients of the higher harmonics in Eq.~(\ref{pd})
can be written in terms of $ \phi^{(1)}_0$. For instance, for $\alpha=3$, we have:
$$
\phi^{(3)}_1= \frac{e^{2 \ri \kappa}+e^{\ri \kappa} }{1 - e^{\ri \kappa}} u^{(1)}_1\phi^{(1)}_0, \label{3rtr}
$$

By a proper rescaling of $ \phi^{(1)}_0$ and $\mu_1$ Eq.~(\ref{hg}) is equivalent to the standard Zakharov-Shabat
spectral problem of the integrable NLS \cite{zs}.

\section{Multiscale expansion of the first two generalized symmetries} \label{gs}

Lie symmetries of a lattice equation $\mathbb{D}(u_{n,m}, T^{\pm}_n u_{n,m}, T^{\pm}_m u_{n,m},  \ldots )=0$ are given by those continuous
transformations which leave the equation invariant. From the infinitesimal point of view 
they are obtained by requiring the infinitesimal invariant condition 
\beq \label{cca2}
\left. {\rm pr} \, \widehat X_{n,m} \, \mathbb{D}  \, \right|_{\mathbb{D} =0} =0,
\eeq
where 
\beq \label{ccb2}
 \widehat X_{n,m} = F_{n,m} ( u_{n,m}, T^{\pm}_n u_{n,m}, T^{\pm}_m u_{n,m},  \ldots) \partial_{u_{n,m}}.
 \eeq
By $ {\rm pr} \, \widehat X_{n,m}$ we mean the prolongation of the infinitesimal generator 
$\widehat X_{n,m}$  to all points appearing in $\mathbb{D}=0$.

If $F_{n,m} = F_{n,m}( u_{n,m})$ then we get {\it point symmetries} and the procedure to get them from Eq.~(\ref{cca2})
is purely algorithmic \cite{lw6}. {\it Generalized symmetries}  are obtained when $F_{n,m} = F_{n,m} ( u_{n,m}, T^{\pm}_n u_{n,m}, T^{\pm}_m u_{n,m},  \ldots)$. 
In the case of nonlinear discrete equations,
the Lie point symmetries are not very common, but, if the equation is integrable and there exists 
a Lax pair, it is possible to construct an infinite family of generalized symmetries. 

In correspondence with the infinitesimal generator (\ref{ccb2}) 
we can in principle construct 
a group transformation by integrating the initial boundary problem
\beq \label{s1}
\frac{d \tilde u_{n,m}(\lambda)}{d \lambda} = 
F_{n,m} ( \tilde u_{n,m} (\lambda) , T^{\pm}_n \tilde u_{n,m} (\lambda), T^{\pm}_m \tilde u_{n,m} (\lambda),  \ldots),
\qquad \tilde u_{n,m}(\lambda =0) = u_{n,m},
\eeq
where   $\lambda \in \mathbb{R}$ is the
continuous Lie group parameter.
This can be done effectively only in the case of point symmetries, as in the
generalized 
case we have a differential-difference equation for which we cannot find the  solution for a generic initial data, but, at most, we can find some particular solutions. 
 Eq.~(\ref{cca2}) is equivalent to the request that the $\lambda$-derivative of the equation $\mathbb{D}=0$, 
written for $\tilde u_{n,m}(\lambda)$, 
is identically satisfied when the $\lambda$-evolution of $\tilde u_{n,m}(\lambda)$ is given by Eq.~(\ref{s1}).
This is also equivalent to say  that the flows (in the group parameter space) given by Eq.~(\ref{s1}) 
are compatible or commute with $\mathbb{D}=0$.

In \cite{lp_sym} one can find an infinite hierarchy of integrable generalized symmetries for the lpKdV 
(\ref{kdv}) constructed by looking at  the isospectral deformations of the Lax pair. The first two symmetries of 
this hierarchy are given by
\begin{gather}
 \frac{d \tilde u_{n,m}}{ d \lambda}= \frac{1}{2p + ( T^-_n - T_n ) \tilde u_{n,m}} -\frac{1}{2p}, \label{simm1} \\
\frac{d \tilde u_{n,m}}{ d \lambda}=
 \frac{1}{[2p + ( T^-_n - T_n ) \tilde u_{n,m}]^2} \left[  \frac{1}{2p + ( 1 - T^2_n ) \tilde u_{n,m} } +
 \frac{1}{2p + ( T^{-2}_n - 1 ) \tilde u_{n,m}}\right]-\frac{1}{4p^3} .\label{simm2} 
\end{gather}
The constant terms appearing in the r.h.s.~of Eqs.~(\ref{simm1},\ref{simm2}) 
ensure that the above flows  go asymptotically to zero
as $\tilde u_{n,m} \rightarrow {\rm{cost}}$. 

To perform the multiscale expansion of the generalized symmetries (\ref{simm1},\ref{simm2})  we consider the following development
for the field $\tilde u_{n,m}$, see Eq.~(\ref{lk}):
\beq 
\label{12}
\tilde u_{n,m} =\sum_{\alpha\in\mathbb{Z}}\sum_{k=1}^{\infty} \frac{1}{N^k}
\tilde u^{(\alpha)}_k (n_2, \{m_i\}_{i=2}^K,\{\lambda_i\}_{i=0}^{K^\prime})
e^{ \ri \alpha(\kappa
n-\omega m)}, \qquad  \qquad \tilde u_k^{(-\alpha)}=\bar { \tilde u}_k^{(\alpha)},
\eeq
where $\lambda_i = \lambda/N^i$ are the slow-varying group parameters, $n_2$ is given by Eq. (\ref{solu})
and $\tilde u_{n,m} (\{\lambda_i=0\}_{i=0}^{K^\prime})= u_{n,m}$.

Since Eq.~(\ref{nlsr}) involves the harmonic $u_1^{(1)}$ we are actually interested just in those
equations, arising from the multiscale expansions of the symmetries (\ref{simm1}, \ref{simm2}),
which are written in terms this harmonic. The following statement holds.

\begin{theorem}
The multiscale expansion up to order $1/N^4$ of the symmetry (\ref{simm1}) 
gives the following symmetries for the dNLS  (\ref{nlsr}) (after a reparametrization
of the group parameters):
\bea
&&\mathcal{O}(1/N): \quad  \; \, \frac{\partial \tilde u_1^{(1)}}{\partial  \lambda} = \ri \tilde u_1^{(1)}, \label{h1}\\
&&\mathcal{O}(1/N^2): \quad 
 \frac{\partial  \tilde u_1^{(1)}}{\partial  \lambda_1}=  
\delta_{n_2} \tilde u_1^{(1) } ,  \label{h2}\\
&&\mathcal{O}(1/N^3): \quad 
\frac{\partial  \tilde u_1^{(1)}}{\partial  \lambda_2}=\delta_{m_2} \tilde u_1^{(1)},  \label{h3} \\
&&\mathcal{O}(1/N^4): \quad 
\frac{\partial  \tilde u_1^{(1)}}{\partial  \lambda_3}= \rho_1
\delta_{n_2}^3 \tilde u_1^{(1)} + 3 \rho_2
 |\tilde u_1^{(1)}|^2 \delta_{n_2}
  \tilde u_1^{(1)},\label{h4}
\eea
with initial condition $ \tilde u_1^{(1)} (\lambda=0,\lambda_1=0,\lambda_2=0,\lambda_3=0)= u_1^{(1)} $. Eqs. 
(\ref{h1},\ref{h2},\ref{h3}) provide point symmetries of Eq.~(\ref{nlsr}), while Eq.~(\ref{h4}) is a generalized symmetry of 
 Eq.~(\ref{nlsr}).
\end{theorem}

\begin{proof} 
The proof is done by a direct computation by taking into account the results contained in 
Theorem \ref{th1}.

Inserting Eq.~(\ref{12}) in the first symmetry (\ref{simm1}) we get the following determing equations:
\bea
&& \mathcal{O}(1/N): \quad  \; \, \frac{ \partial \tilde u_1^{(1)}}{ \partial \lambda} = \frac{\ri}{2 p^2} \sin \kappa \,  \tilde u_1^{(1)}, \label{simm1a}  \\
&& \mathcal{O}(1/N^2): \quad  \frac{\partial \tilde u_2^{(1)}}{\partial \lambda}+\frac{\partial \tilde u_1^{(1)}}{\partial \lambda_1} = \frac{\ri}{2
  p^2} 
\left( \sin \kappa \, \tilde u_2^{(1)} - \ri M_1  \cos \kappa \, \delta_{n_2} \tilde  u_1^{(1)}  \right),  \label{simm1b}
\eea
\bea
&&\mathcal{O}(1/N^3): \quad  \frac{\partial \tilde u_3^{(1)}}{\partial \lambda} + \frac{\partial \tilde  u_2^{(1)}}{\partial  \lambda_1}+\frac{\partial 
 \tilde  u_1^{(1)}}{\partial \lambda_2} =  \label{simm1c} \\
&&\qquad \qquad \qquad = \frac{\ri}{2 p^2} \left(  \sin \kappa \, \tilde u_3^{(1)}  - \ri M_1 \cos \kappa \,  \delta_{n_2} \tilde u_2^{(1)}  
+ \frac{M_1^2}{2}  \sin \kappa \,  \delta_{n_2}^2\tilde  u_1^{(1) } \right) +  \nonumber \\
&&  \qquad \qquad \qquad + \frac{\ri }{p^3} \left( - \ri \sin\kappa \,  \sin (2 \kappa) \bar {\tilde  u}_1^{(1)}
  u_2^{(2)} + 
 M_1 \sin \kappa \, \tilde u_1^{(1)} \delta_{n_2} \tilde u_1^{(0)} \right) +  \nonumber \\
 &&\qquad \qquad \qquad +\frac{3 \ri }{2
p^4 } \sin^3 \kappa \,  |\tilde u_1^{(1)}|^2
  \tilde u_1^{(1)}, \nonumber 
  \eea
\bea
&&\mathcal{O}(1/N^4): \quad  \frac{\partial \tilde u_{4}^{(1)}} {\partial\lambda}+ \frac{\partial \tilde u_{3}^{(1)}} {\partial\lambda_{1}} + 
\frac{\partial \tilde u_{2}^{(1)}} {\partial\lambda_{2}}+ \frac{\partial \tilde u_{1}^{(1)}} {\partial\lambda_{3}} = \label{simm1d}
\\
&&\qquad \qquad \qquad = \frac{\ri}{2 p^2} \left(  \sin \kappa \,  \tilde u_{4}^{(1)}  -\ri M_{1}\cos \kappa \,  \delta_{n_2}\tilde u_{3}^{(1)} + 
\frac{M_1^2}{2}  \sin \kappa \,  \delta_{n_2}^2 \tilde u_1^{(1) } - \right.  \nonumber \\
&&\qquad \qquad \qquad  \qquad \qquad \left. - \frac{\ri M_1^3}{3} \cos \kappa \, \delta_{n_2}^3\tilde u_{1}^{(1)}
\right)+  \nonumber \\
&&\qquad \qquad \qquad  + \frac{\ri }{p^3} \left[ -\ri \sin \kappa \,  \sin (2 \kappa)  \,  \left( \bar {\tilde u}_1^{(1)}
  \tilde u_3^{(2)} + \bar {\tilde u}_2^{(1)}
  \tilde u_2^{(2)} \right) + \right.  \nonumber \\
&&\qquad \qquad \qquad   \qquad \quad  \left. + M_1 \sin \kappa \, \left( \bar {\tilde u}_1^{(1)} \delta_{n_2} \tilde u_2^{(2)}  + 
\tilde u_1^{(1)} \delta_{n_2} \tilde u_2^{(0)} +
 \tilde u_2^{(1)} \delta_{n_2} \tilde u_1^{(0)} \right) - \right.  \nonumber \\
 &&\qquad \qquad \qquad  \qquad \quad \left. - \ri M_1^2 \cos \kappa \,  \tilde u_{1}^{(1)} \delta_{n_2}^2 \tilde u_{1}^{(0)} \right] +  \nonumber  \\
&&\qquad \qquad \qquad + \frac{3 \ri } {2p^{\,4}} \left[
- \ri M_{1}\cos \kappa \sin^2 \kappa \, (\tilde u_{1}^{(1)})^2 \delta_{n_2} \bar {\tilde u}_{1}^{(1)}+ \right. \nonumber \\
&& \qquad \qquad \qquad  \qquad \quad \quad + \left. \sin^3 \kappa
\left( \bar {\tilde u}_{2}^{(1)}(\tilde u_{1}^{(1)})^2+2\tilde u_{2}^{(1)}|\tilde u_{1}^{(1)}|^2\right) \right]. \nonumber 
\eea

Let us consider Eq.~(\ref{simm1a}); by the reparametrization $\lambda \mapsto 2 p^2\lambda/\sin \kappa$,
  Eq.~(\ref{simm1a}) is equivalent to Eq.~(\ref{h1}). This is the first point symmetry of the
dNLS ~(\ref{nlsr}) and it corresponds to a phase symmetry.

Eq.~(\ref{simm1b}) has to be split into the following equations to avoid secularities:
\bea
&& \frac{\partial \tilde u_1^{(1)}}{\partial  \lambda_1}=  \frac{M_1}{2 p^2}\cos \kappa \, 
\delta_{n_2} \tilde u_1^{(1) } , \label{r1}\\
 && \frac{ \partial \tilde u_2^{(1)}}{\partial  \lambda} = \frac{\ri}{2 p^2} \sin \kappa \,  \tilde u_2^{(1)}. \label{r2}
\eea

From Eq.~(\ref{r2}) we see that $\tilde u_2^{(1)}$ depends on $\lambda$ as $\tilde u_1^{(1)}$. 
Eq.~(\ref{r1}) provides the second point symmetry  (\ref{h2}) of the dNLS  (\ref{nlsr}),
corresponding to translations w.r.t.~the index $n_2$,  after the reparametrization 
$\lambda_1 \mapsto 2 p^2\lambda_1/ (M_1 \cos \kappa$). 

From Eq.~(\ref{simm1c}), taking into
account Eqs. (\ref{simm1a},\ref{r2}) and the secularity conditions, a straightforward algebra 
and the reparametrization $\lambda_2 \mapsto 4 p^2 \rho_1 \lambda_2 /(M_1^2 \sin \kappa)$
leads to
$$
\ri \frac{\partial  \tilde u_1^{(1)}}{\partial  \lambda_2}= \rho_1 \delta_{n_2}^2 \tilde u_1^{(1)}+ \rho_2  \tilde u_1^{(1)} |\tilde u_1^{(1)}|^2,
$$
which leads to Eq.~(\ref{h3}) thanks to Eq.~(\ref{nlsr}). Eq.~(\ref{h3}) means that the dNLS ~(\ref{nlsr}) is invariant under
translations w.r.t.~the index $m_2$.

Finally, Eq.~(\ref{simm1d}) gives Eq.~(\ref{h4}) after a long computation by taking into account
 Eqs. (\ref{simm1a},\ref{simm1b},\ref{simm1c}). In this last case the reparametrization
of the group parameter reads  $\lambda_3 \mapsto 12 p^2 \rho_1 \lambda_3  / (M_1^3 \cos \kappa)$.

\end{proof}

A computation up to order $1/N^4$, similar to the one just done for the symmetry (\ref{simm1}), shows that the multiscale expansion of the
second generalized symmetry (\ref{simm2}) of the lpKdV  (\ref{kdv}) gives the same symmetries 
(\ref{h1},\ref{h2},\ref{h3},\ref{h4}), after suitable reparametrizations of the group parameters.

\section{Concluding remarks} \label{c}
In this paper we have considered the multiscale expansion of the spectral problem and of the symmetries of the partial difference integrable lattice potential KdV equation.  By a proper choice of the spectral problem of the lpKdV we have been able to derive from it the spectral problem of the reduced equation, a nonlinear Schr\"odinger equation. We then did the multiscale expansion of two generalized symmetries. A generalized symmetry provides us with the point and generalize symmetries of the nonlinear Schr\"odinger equation. At each order of the multiscale approximation, we get by reduction from the request that no secular condition exists, a higher order symmetry. The same calculation for other generalized symmetries do not provide anything new. All the information concerning the whole hierarchy of generalized symmetries for the NLS is contained in the first generalized symmetry for the lpKdV.

\section*{Acknowledgments}

MP was partially supported by
the European Community through the FP6 Marie Curie RTN ENIGMA (contract number
MRTN-CT-2004-5652).
DL,  MP and CS were partially supported by the PRIN project 
``Metodi geometrici nella teoria delle onde non lineari ed applicazioni, 2006" of the Italian Minister
for Education and Scientific Research.
RHH was partially supported by the Region of Madrid and Universidad
Polit\'ecnica de Madrid (UPM) through the grant ref.~CCG06-UPM/
MTM-539 and the Spanish Ministry of Science project MTM2006-13000-
C03-02.



\begin{thebibliography}{99}
\small

\bibitem{Ag}
\textsc{Agrotis M}, \textsc{Lafortune S} and \textsc{Kevrekidis P\,G},
On a discrete version of the KdV equation,
{\it Discr. Cont. Dyn. Sist.} {\bf 2005} supp. 22--29.

\bibitem{boiti}
\textsc{Boiti M}, \textsc{Bruschi M}, \textsc{Pempinelli F} and \textsc{Prinari F},
 A discrete Schršdinger spectral problem and associated evolution equations,
{\it J. Phys. A: Math. Gen.} {\bf 36}  (2003) 139--149

\bibitem{ce}
\textsc{Calogero F} and \textsc{Eckhaus W},
Nonlinear evolution equations, rescalings, model PDEs and their integrability. I,
{\it Inv. Prob.} {\bf 3} 2 (1987) 229--262.

\textsc{Calogero F} and \textsc{Eckhaus W},
Nonlinear evolution equations, rescalings, model PDEs and their integrability. II,
{\it Inv. Prob.} {\bf 4} 1 (1987) 11--33.

\bibitem{dms}
\textsc{Degasperis A}, \textsc{Manakov S V} and  \textsc{Santini P M}, 
Multiple-scale perturbation beyond the nonlinear Schroedinger equation. I,
{\it Phys. D} {\bf 100} (1997) 187--211.


\bibitem{dp}
\textsc{Degasperis A} and \textsc{Procesi D}, 
Asymptotic Integrability, in {\it Symmetry and Perturbation Theory SPT98} (1999) 23--37.

\bibitem{mf}
\textsc{Flaschka H}, 
The Toda lattice. I. Existence of integrals,
{\it Phys. Rev. B} {\bf  9}  (1974) 1924--1925. 

\textsc{Manakov S\,V},
Complete integrability and stochastization in discrete dynamic systems,
{\it Zhur. Eksp. i Teor. Fiziki.} {\bf 67} (1974) 543--555. 

\bibitem{HLPS}
\textsc{Hernandez Heredero R}, \textsc{Levi D}, \textsc{Petrera M} and \textsc{Scimiterna C}, 
Multiscale expansion of the lattice potential KdV equation on functions of an infinite slow-varyness order,
 {\it Journ. Phys. A} {\bf 40} (2007), F831--F840.

\bibitem{HLPS_t} 
\textsc{Hernandez Heredero R}, \textsc{Levi D}, \textsc{Petrera M} and \textsc{Scimiterna C}, 
Multiscale expansion on the lattice and integrability or linearizability of partial difference equations, in preparation.

\bibitem{Jordan} 
\textsc{Jordan C}, 
Calculus of finite differences, 
R\"ottig and Romwalter, Sopron, 1939.

\bibitem{km}
\textsc{Kodama Y} and \textsc{Mikhailov A\,V}, 
Obstacles to asymptotic integrability, in {\it Algebraic Aspects
of Integrable Systems, in Memory of Irene Dorfman}, Progress in Nonlinear Differential
Equations, Vol. 26,  Birkh\"auser, Boston, 1996, 173--204.

\bibitem{lm}
\textsc{Leon J} and \textsc{Manna M},
Multiscale analysis of discrete nonlinear evolution equations,
 {\it Journ. Phys. A} {\bf 32} (1999) 2845--2869.

\bibitem{levi}
\textsc{Levi D}, 
Multiple-scale analysis of discrete nonlinear partial difference equations: the reduction of the lattice potential KdV,
{\it Jour. Phys. A} {\bf 38} (2005) 7677--7685.

\bibitem{LevHer}
\textsc{Levi D} and \textsc{Hernandez Heredero R},
Multiscale analysis of discrete nonlinear evolution equations: the Reduction of the dNLS,
{\it Jour. Nonlinear Math. Phys.} {\bf 12} 1 (2005) 440--455.

\bibitem{lp}
\textsc{Levi D} and \textsc{Petrera M}, 
Discrete reductive perturbation technique,
{\it Jour. Math. Phys.} {\bf 47} (2006) 043509.

\bibitem{lp_sym} 
\textsc{Levi D} and \textsc{Petrera M}, 
Continuous symmetries of the lattice potential KdV equation,
{\it Jour. Phys. A} {\bf 40} (2007) 4141--4159.

\bibitem{lw6}
\textsc{Levi D} and \textsc{Winternitz P},
Continuous symmetries of difference equations,
{\it Jour. Phys. A} {\bf  39} (2006)  R1--R63.

\bibitem{frank}
\textsc{Nijhoff F\,W} and \textsc{Capel H\,W},
The discrete Korteweg-de Vries equation,
{\it Acta Appl. Math.}  {\bf 39}  (1995) 133--158.

\bibitem{zs}
\textsc{Novikov S\,P}, \textsc{Manakov S\,V}, \textsc{Pitaevski L\,P} and \textsc{Zakharov V\,E},
Theory of solitons: the inverse scattering method,  Elsevier, New York, 1984.

\bibitem{t1}
\textsc{Taniuti T},
Reductive perturbation method for nonlinear wave propagation,
{\it Prog. Theor. Phys.} {\bf 55 }  (1974) 1654--1676.


\bibitem{t2}
\textsc{Taniuti T}  and \textsc{Nishihara K},
Nonlinear waves,  Pitman, Boston, 1983.

\bibitem{KZ}
\textsc{Zakharov V\,E}  and \textsc{Kuznetsov E\,A},
Multi-scale expansions in the theory of systems integrable by the
inverse scattering transform,
{\it Phys. D}  {\bf 18}  (1986) 455--463.

\end{thebibliography}
\end{document}